\documentclass[aps,prl,twocolumn,groupedaddress]{revtex4}
\usepackage{graphicx}
\newcommand{\be}{\begin{equation}}
\newcommand{\ee}{\end{equation}}
\newcommand{\bd}{\begin{displaymath}}
\newcommand{\ed}{\end{displaymath}}

\begin{document}
\title{Microscopic measurement of photon echo formation in groups of individual excitonic transitions}
%
%
\author{Wolfgang Langbein}
\affiliation{Department of Physics and Astronomy, Cardiff
University, Cardiff CF24 3YB, United Kingdom}
\email[]{LangbeinWW@cardiff.ac.uk}
\author{Brian Patton}
\affiliation{Experimentelle Physik IIb, Universit\"{a}t Dortmund,
Otto-Hahn-Str. 4, 44221 Dortmund, Germany}

\begin{abstract}
The third-order polarization emitted from groups of individual
localized excitonic transitions after pulsed optical excitation is
measured. We observe the evolution of the nonlinear response from
the case of a free polarization decay for a single transition, to
that of a photon echo for many transitions. The echo is shown to
arise from the mutual rephasing of the emission from individual
transitions.

\end{abstract}


\date{\today}
\maketitle

Nonlinear optical spectroscopy is a powerful technique when investigating the electronic structure and
dynamics of matter. Specifically, the third-order nonlinearity probed in four-wave mixing (FWM),
spectral hole-burning, and pump-probe experiments can be used to determine homogeneous line-shapes
of inhomogeneously broadened transitions. This was first employed for nuclear spin transitions
excited by radio-frequencies, which show spin echoes \cite{HahnPR50}. The availability of ultrafast
laser pulses allowed the echo technique to be extended to optical frequencies
\cite{KurnitaPRL64,ChoJCP92}. More recent experiments have seen the effect of inhomogeneous
broadening be circumvented by measuring on individual systems, such as single atoms in traps,
single molecules on surfaces, or single localized exciton states in semiconductor quantum dots.
This is currently only possible in the optical frequency range in which highly sensitive detectors
are available. In such experiments, the measurement time has to be much longer than the dephasing
time in order to collect a sufficient number of photons. During this time, fluctuations in the
environment lead to slow spectral diffusion of the frequency of the investigated transitions (e.g.
by the Stark effect), resulting in an inhomogeneous broadening in the measurement due to the
time-ensemble \cite{MukamelBook99}. As a result, the typical line-width of semiconductor quantum
dots at low temperatures measured in single-dot photoluminescence is 10-1000\,$\mu$eV
\cite{BayerPRB02,LeossonPSS00,TurckPRB00, GindelePRB99a}, while the homogeneous line-width
determined in photon echo experiments is about 1\,$\mu$eV
\cite{KuribayashiPRB98,BorriPRL01,LangbeinPRB04a} and can be given by the natural line-width, i.e.
the radiative decay rate. FWM spectroscopy on individual exciton transitions can distinguish
between these two broadening mechanisms. However, until now FWM on single exciton states was only
performed using continuous wave excitation with non-degenerate frequencies
\cite{GuestScience01,BonadeoPRL98}. Therefore, the formation of the photon echo from the
interference of the FWM polarizations of individual transitions has, up to now, not been observed.

In this work, we present transient four-wave mixing measurements on
individual, localized excitons. Using a multichannel heterodyne
detection, the frequency-resolved third-order polarization is
measured in amplitude and phase, so that both the time and the
spectrally resolved third-order polarization can be retrieved. Analogous to
the introduction of pulsed nuclear magnetic resonance spectroscopy
to replace scanning techniques, the detection sensitivity is
increased significantly by the multichannel detection. This enables
us to investigate the evolution of the nonlinear response from the case of a
free polarization decay to that of a photon echo as the number of
individual transitions in an inhomogeneously broadened ensemble is increased.

The size of the investigated excitonic states is much smaller than the wavelength of the resonant
light. Therefore, the emitted polarization of an individual state is essentially isotropic (apart
from polarization effects), rendering the commonly used directional selection of the four-wave
mixing signal useless. To discriminate the four-wave mixing signal in such a case, we have to use
the phase coherence of the signal relative to the excitation pulses, which is determined by the
form of the third-order polarization $P^{(3)}\propto E_1^* E_2 E_2$, with the excitation electric
fields $E_{1,2}$. In order to be sensitive to the phase of the emitted FWM field, we detect it via
its interference with a reference field $E_{\rm r}$. Such a detection principle was used to measure
pump-probe and four-wave mixing in waveguides \cite{HallOL92,BorriOC99} and in planar InAs quantum
dot ensembles \cite{LangbeinPRB04a}. In these investigations of large ensembles of excitonic states
($N>10^5$), the FWM was emitted as a photon echo \cite{BorriPRL01} of $\approx 100$\,fs duration
given by the inverse inhomogeneous broadening of the transition energies in the excited ensemble.
Conversely, when probing individual transitions, the signal of each transition is expected to be
emitted as a free polarization decay, with a decay time given by the intrinsic dephasing time of
the transition, which can be many orders of magnitude longer than the ensemble photon echo. In
order to efficiently measure a signal of such rich spectral complexity, a multichannel detection is
needed. We employ a multichannel heterodyne scheme using spectrally resolved detection with a
charge-coupled device (CCD), and retrieve the signal by spectral interferometry
\cite{LepetitJOSAB95}. This scheme provides a simultaneous measurement of all spectral components
of the signal in both amplitude and phase, allowing the determination of the signal in both
frequency- and time-domain by Fourier-transform.

\begin{figure}
\includegraphics*[width=8cm]{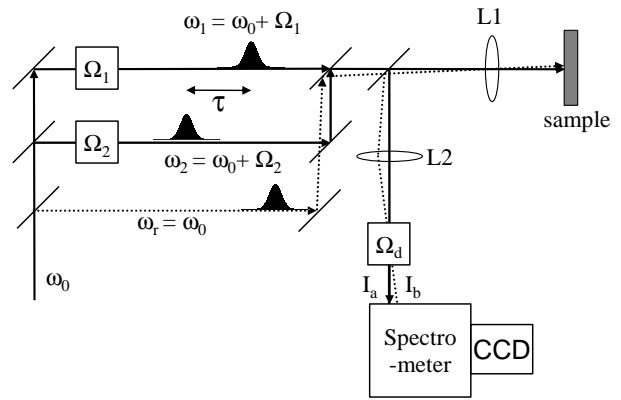}
\caption[]{ \label{fig:setup} Scheme of the experimental setup. Boxes:
Acousto-optical modulators of the indicated frequency. L1: High numerical
aperture (0.85) microscope objective. Spectrometer: Imaging spectrometer of
15\,$\mu$eV resolution.}\end{figure}

A schematic overview of the experimental setup is given in
Fig.\,\ref{fig:setup}. We start with optical excitation pulses of 0.2\,ps
duration at 76\,MHz repetition rate from a mode-locked Ti:sapphire laser with a
center frequency $\omega_0$. Two pulse trains 1,2 are created by a
beam-splitter, frequency shifted with acousto-optical modulators (AOMs) by the
radio frequencies $\Omega_1/2\pi=79$\,MHz and $\Omega_2/2\pi=80$\,MHz, and
recombined into the same spatial mode, with a relative delay time $\tau$,
positive for pulse 1 leading. A frequency-unshifted reference pulse train is
recombined with the pulse trains 1,2
 in such a way as to pass through the same optics but in a sightly different direction. In this manner, passive phase stability between the excitation and
the reference beams is achieved over the whole optical path. The pulses are
focussed onto the sample using a microscope objective of a numerical aperture
NA=0.85 mounted in a helium bath cryostat. The FWM signal is collected by the
same objective, and directed into a mixing AOM, in such a way that the
diffracted beam of the signal overlaps with the reference beam reflected by the
sample and vice versa. The mixed beams a,b (see Fig.\,\ref{fig:setup}) are
spectrally resolved and detected by a liquid nitrogen cooled silicon CCD,
measuring their spectrally and time-resolved intensities $I_{\rm
a,b}(\omega,t)$. In this notation, the variable $\omega$ describes the optical
frequencies, with a resolution given by the spectrometer ($\approx$3\,GHz),
while the time variable $t$ describes the low-frequency dynamics, which
contains the modulation due to the frequency shifts by the AOMs \footnote{This
separation of time-scales is possible for dephasing times shorter than the
radio frequency period and the pulse repetition time.}. The mixing AOM
down-shifts the signal field $E_{\rm s}$ in frequency by $\Omega_{\rm d}$ when
deflected into the reference field $E_{\rm r}$ , while the reference is
frequency up-shifted by $\Omega_{\rm d}$ when deflected into the signal. In
this way, the interference frequency of signal and reference fields in the
beams a,b is shifted by $\Omega_{\rm d}$. The deflection efficiency is adjusted
to 50\%, so that the detected intensities are given by $2I_{\rm
a,b}(\omega,t)=|E_{\rm r}|^2 +|E_{\rm s}|^2 \pm 2\Re(E_{\rm r}E_{\rm
s}^*e^{i\Omega_{\rm d} t})$. Since we detect $I_{\rm a,b}(\omega,t)$ temporally
integrated over the exposure time $T$ of the CCD (which is typically
50-1000\,ms), only the interference close to zero frequency ($\Omega<1/T$) is
detected. To recover the interference term, we subtract the two detected
intensities, yielding $I_{\rm d}(\omega)=I_{\rm a}-I_{\rm b}=2\int_0^T
\Re(E_{\rm r}E_{\rm s}^*e^{i\Omega_{\rm d}t})dt$.

Having measured the interference intensity $I_{\rm d}(\omega)$, we deduce the signal field in
amplitude and phase by spectral interferometry \cite{LepetitJOSAB95}, using the fact that the
reference field is adjusted to precede the signal field in time: $F(\Theta(t)F^{-1}(I_{\rm
d}(\omega))) = E^*_{\rm r}(\omega)E_{\rm s}(\omega)e^{i\Omega_{\rm d} t}$ with the Heaviside
function $\Theta(t)$, and the Fourier-transform operator $F$. The time-range of the resulting
time-resolved signal field of this technique is limited by the spectral resolution to about
100\,ps, which is much smaller than the repetition period. The reference field amplitude can be
determined by blocking the signal beam and measuring $I_{\rm a,b}=|E_{\rm r}|^2$ in this case. The
reference phase can be determined by nonlinear pulse characterization techniques, or can be
calculated from the chirp introduced by the optical components in the setup. Knowing the reference
field $E_{\rm r}(\omega)$, the signal field $E_{\rm s}(\omega)$ can be determined. The choice of $\Omega_{\rm d}$ selects the detected interference. For $\Omega_{\rm d}=\Omega_{1,2}$, the reflected
excitation pulses 1,2 are measured, while for $\Omega_{\rm d}=2\Omega_{2}-\Omega_{1}$, the emitted
FWM field $\propto E_1^* E_2 E_2$ is measured. Higher-order non-linearities, like six-wave mixing,
can be detected analogously.

\begin{figure}
\includegraphics*[width=8cm]{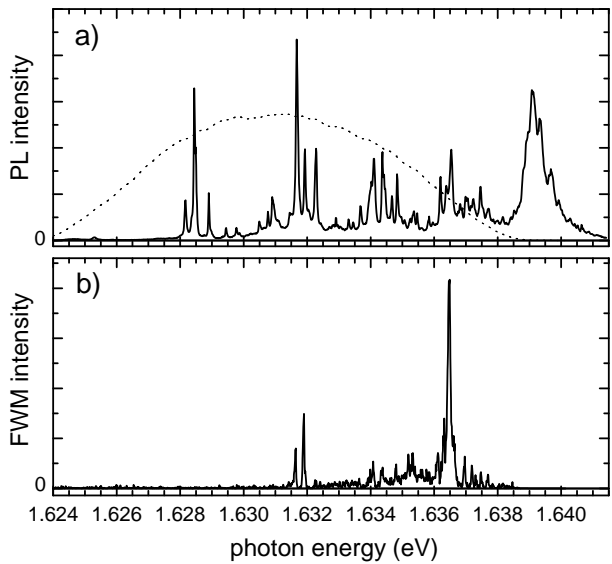}
\caption[]{ \label{fig:plfwm} a) Confocally excited / detected PL
spectrum of a $(0.5\,\mu$m$)^2$ area of a AlAs/GaAs QW with a
thickness of about 20.8\,ML (6\,nm). The spectrum of the excitation
pulses used in the FWM experiment of b) and
Fig.\,\ref{fig:temporalfwm} is shown as dotted line. b) Spectrally
resolved FWM intensity at $\tau=1$\,ps.}\end{figure}

The investigated sample \cite{LeossonPRB00} consists of an MBE--grown AlAs/GaAs/AlAs single quantum
well with a thickness increasing from approximately 4\,nm to 10\,nm over a total lateral size of
200\,mm. The growth was interrupted for 120\,s at each interface, allowing for the formation of
large monolayer islands on the growth surface. The sample was antireflection coated, and was held
in a helium cryostat at a temperature of T=5\,K. Exciting non-resonantly at 1.96\,eV, focussed to
the diffraction limit, the confocally detected photoluminescence (PL) is shown in
Fig.\,\ref{fig:plfwm}a. In order to select individual states within the optical resolution of the
experiment (0.5\,$\mu$m), we adjusted the fractional monolayer thickness of the QW to be about -0.2
ML \cite{SavonaPSS04}, yielding very few localized exciton states in the largest monolayer (ML)
thickness. This was done by monitoring the PL spectrum while moving the excitation spot along the
QW thickness gradient. Individual emission lines are visible in the low-energy part of the
spectrum, corresponding to individual localized excitons. Due to the diffusion of the excited
carriers prior to recombination, the spatial resolution of the PL spectrum is not significantly
improved by the confocal excitation, and is essentially determined by the diffraction limited
intensity resolution of $0.61 \lambda/{\rm NA}$ in the emission imaging. The same region was
investigated by the FWM technique. The laser pulse spectrum shown in Fig.\,\ref{fig:plfwm}a excited
only the excitons localized in the lower monolayer in order to avoid large excited exciton
densities. The measured spectrally resolved FWM at $\tau=1$\,ps is given in Fig.\,\ref{fig:plfwm}b.
It consists of several sharp resonances of 20-30\,$\mu$eV FWHM. The FWM intensity is proportional
to the third power of the excitation intensity,
which improves the spatial resolution to $0.36 \lambda/{\rm NA}\approx 320$\,nm. 
Only at the higher energy side of the PL emission are FWM resonances of significant strength
observed. The FWM intensities of the resonances are not clearly correlated to the PL intensities.
This can be understood considering the properties of the resonances determining the respective
signal strength. The FWM intensity is determined only by the fourth power of the optical transition
dipole moment $\mu$ of the resonance. The PL intensity instead is determined by the radiative rate,
proportional to $\mu^2$, and also by the relaxation dynamics of excitons into the localized state.
Additionally, in the PL we use non-resonant excitation, for which charged exciton emission
\cite{TischlerPRB02}, having a binding energy of about 4\,meV, could be present, explaining the
lower energy peaks (e.g. at 1.628\,eV) in the PL.

\begin{figure}
\includegraphics*[width=8cm]{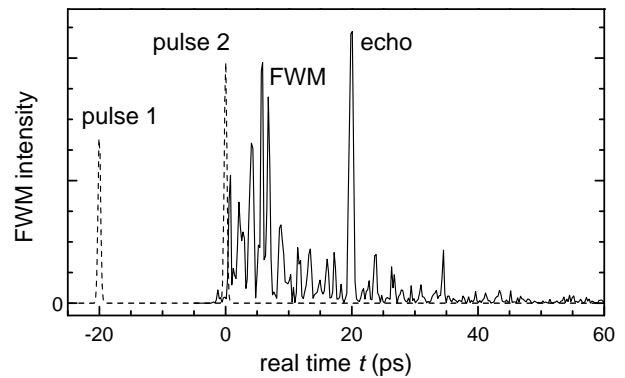}
\caption[]{ \label{fig:temporalfwm} Time-resolved FWM intensity from
a group of exciton states localized within an $(0.5\,\mu$m$)^2$
area. Excitation pulses 1,2 (dashed line) and time-resolved FWM
intensity (solid line) for a delay time of $\tau=20$\,ps are shown.
Excitation energies were 1.0 (1.9) \,fJ for pulse 1(2),
corresponding to an intensity per pulse of 0.4(0.75)\,$\mu$J/cm$^2$
at the quantum well.}\end{figure}

The temporal sequence of the experiment, consisting of the arrival of pulse 1, pulse 2, and the
subsequent emission of the FWM intensity can be measured in the experiment by choosing $\Omega_{\rm
d}$ equal to $\Omega_{1}$, $\Omega_{2}$, or $2\Omega_{2}-\Omega_{1}$, respectively, and is shown in
Fig.\,\ref{fig:temporalfwm}. Pulse 1 arrives at $t=-20$\,ps, and creates a first-order polarization
that oscillates at the resonance frequencies of the system. Upon the arrival of pulse 2 at $t=0$ ,
this polarization interferes with pulse 2 and creates a density that oscillates with the frequency
$\Omega_{2}-\Omega_{1}$. The polarization that is created by pulse 2 and that is proportional to
this density is the third-order polarization. It has the frequency $2\Omega_{2}-\Omega_{1}$, and is
the source of the self-diffracted FWM. The FWM is thus expected to start at $t=0$, in agreement
with the experimental observation. Since several resonances of frequencies $\omega_k$ are emitting,
their superposition leads to a strong temporal beating in the signal. Generally, we can express the
emitted FWM field $E^{(3)}$ for $\tau>0$ for a set of $N$ two-level systems as:

\be \label{eqn:fwm} E^{(3)}(t,\tau)=\sum_{k=1}^N
\mu_k^4E_1(\omega_k)E_2^2(\omega_k)e^{i\omega_k(t-\tau)-\gamma_k(t+\tau)}.
\ee

\begin{figure}[b]
\includegraphics*[width=8cm]{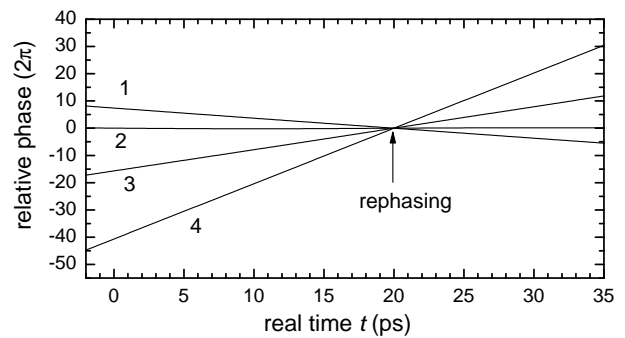}
\caption[]{ \label{fig:echophase} Measured relative phase evolution of the FWM field of 4
individual transitions of an ensemble. The corresponding FWM spectrum is shown in the middle row of
Fig.\,\ref{fig:echoensemble}. The phase noise of about 0.3 rad is not readily discernible due to
the large phase scale.}\end{figure}
\begin{figure}[t]
\includegraphics*[width=8cm]{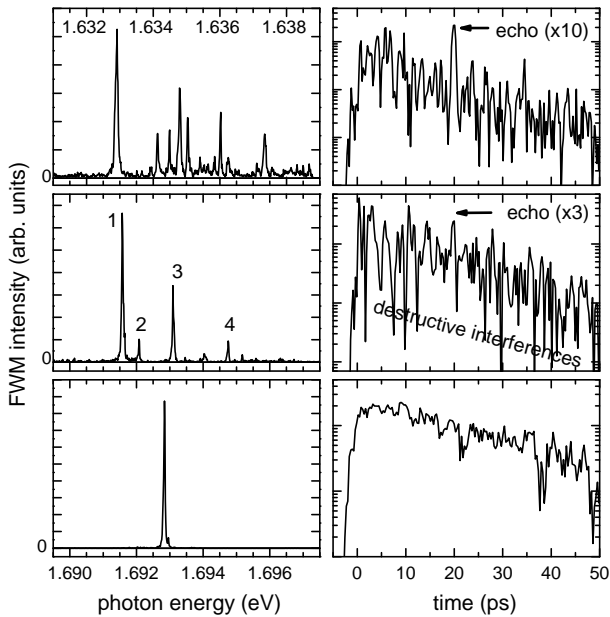}
\caption[]{ \label{fig:echoensemble} FWM intensity for
$\tau=20$\,ps, spectrally resolved (left) and time-resolved (right),
for exciton state ensembles of different size. The estimated
fractional enhancement by the superradiance at the photon echo time
$t=20$\,ps is indicated.}\end{figure}


with the dipole moments of the transitions $\mu_k$ and the
appropriately normalized excitation field amplitudes
$E_{1,2}(\omega)$ of pulse 1,2 at the frequency $\omega$. Here we
assume that the pulse spectral width is much larger than the
dephasing rates $\gamma_k$ of the transitions. At $t=\tau$, the FWM
fields from all two-level systems are predicted to be in phase, and
therefore to interfere constructively. This results in a signal
amplitude $N$ times larger than the individual FWM amplitudes of a
single transition. For other times, the phases are, in general,
randomly distributed due to the distribution of the transition
frequencies, and the enhancement is reduced to $\approx\sqrt{N}$. In
the limit of a large number of systems in the ensemble, the signal
at $t=\tau$ is thus far larger than at other times, and is called a
photon echo. Until now, the signal away from the echo time, in the
random interference time, was not observed. Here, we can see
(Fig.\,\ref{fig:temporalfwm}) the formation of the photon echo by
the constructive interference at $t=20$\,ps. Since we measure the
phase and amplitude of the signal, we can determine the phase
evolution for each resonance separately by applying a spectral
filter. The result of such an analysis for 4 transitions of an
ensemble is shown in Fig.\,\ref{fig:echophase} (the corresponding
spectrally resolved FWM is given in the middle row of
Fig.\,\ref{fig:echoensemble}). In the measured phase evolution, the
rephasing of the FWM fields from the individual transitions at the
photon echo time $t=20$\,ps is directly observed.

Furthermore, the fact that we can determine, in the spectral domain, the number of states
contributing to the time-domain FWM signal allows us, by appropriate selection of regions of the
sample, to control the size of the state ensemble probed in the FWM. By systematically increasing
the number of participating transitions from one to many, we can therefore follow the formation of
the photon echo with increasing ensemble sizes. This evolution is shown in
Fig.\,\ref{fig:echoensemble} for ensembles of about 1, 4, and 10 transitions. With increasing
ensemble size $N$, the intensity enhancement in the photon echo increases roughly $\propto N$, as
expected from eqn.\,(\ref{eqn:fwm}). The peaks off the photon echo time are due to subsidiary
constructive interferences of subgroups of states, and are prominent due to the small number of
participating states.

In conclusion, we have measured the transient third-order polarization of individual quantum states
in both amplitude and phase.  The formation of the photon echo with increasing ensemble size was
experimentally demonstrated and the signal was shown to vary from a free polarization decay for the
single emitter to a photon echo in the ensemble limit; we have thus demonstrated experimentally the
microscopic origin of the photon echo in transient four-wave mixing in a solid state system.

The sample was grown at III-V Nanolab, a joint laboratory between
Research Center COM and the Niels Bohr Institute, Copenhagen
University. Continued support by U.~Woggon is acknowledged. This
work was funded by the German Science Foundation (DFG) within the
Grant WO477/14.

%

\begin{thebibliography}{10}

\bibitem{HahnPR50}
E. Hahn, Phys. Rev. {\bf 80},  580  (1950).

\bibitem{KurnitaPRL64}
N.~A. Kurnita, I.~D. Abella, and S.~R. Hartmann, Phys. Rev. Lett. {\bf 13},
  567  (1964).

\bibitem{ChoJCP92}
M. Cho, N.~F. Scherer, G.~R. Fleming, and S. Mukamel, J. Chem. Phys. {\bf 96},
  5618  (1992).

\bibitem{MukamelBook99}
S. Mukamel, {\em Principles of Nonlinear Optical Spectroscopy} (Oxford, USA,
  1999).

\bibitem{BayerPRB02}
M. Bayer and A. Forchel, Phys. Rev. B {\bf 65},  041308(R)  (2002).

\bibitem{LeossonPSS00}
K. Leosson, J.~R. Jensen, J.~M. Hvam, and W. Langbein, phys. stat. sol. (b)
  {\bf 221},  49  (2000).

\bibitem{GindelePRB99a}
F. Gindele, K. Hild, W. Langbein, and U. Woggon, Phys. Rev. B {\bf 60},  R2157
  (1999).

\bibitem{TurckPRB00}
V. T{\"u}rck {\it et~al.}, Phys. Rev. B {\bf 61},  9944  (2000).

\bibitem{BorriPRL01}
P. Borri {\it et~al.}, Phys. Rev. Lett. {\bf 87},  157401  (2001).

\bibitem{LangbeinPRB04a}
W. Langbein {\it et~al.}, Phys. Rev. B {\bf 70},  033301  (2004).

\bibitem{KuribayashiPRB98}
R. Kuribayashi {\it et~al.}, Phys. Rev. B {\bf 57},  R15084  (1998).

\bibitem{GuestScience01}
J. Guest {\it et~al.}, Science {\bf 293},  2224  (2001).

\bibitem{BonadeoPRL98}
N.~H. Bonadeo {\it et~al.}, Phys. Rev. Lett. {\bf 81},  2759  (1998).

\bibitem{HallOL92}
K.~L. Hall, G. Lenz, E.~P. Ippen, and G. Raybon, Optics Letters {\bf 17},  874
  (1992).

\bibitem{BorriOC99}
P. Borri {\it et~al.}, Optics Commun. {\bf 164},  51  (1999).

\bibitem{LepetitJOSAB95}
L. Lepetit, G. Ch{\'e}riaux, and M. Joffre, J. Opt. Soc. Am. B {\bf 12},  2467
  (1995).

\bibitem{LeossonPRB00}
K. Leosson, J.~R. Jensen, W. Langbein, and J.~M. Hvam, Phys. Rev. B {\bf 61},
  10322  (2000).

\bibitem{SavonaPSS04}
V. Savona, W. Langbein, and G. Kocherscheidt, phys. stat. sol. (c) {\bf 1},
  501  (2004).

\bibitem{TischlerPRB02}
J.~G. Tischler, A.~S. Bracker, D. Gammon, and D. Park, Phys. Rev. B {\bf 66},
  081310(R)  (2002).

\end{thebibliography}
%

\end{document}